\documentclass[%
 aip,
 jmp,%
 amsmath,amssymb,
 reprint,%
]{revtex4-1}

\usepackage{graphicx}
\usepackage{dcolumn}
\usepackage{bm}
\renewcommand{\vec}[1]{\mathbf{#1}}

\begin{document}


\title{Tuning structure and mobility of solvation shells surrounding tracer additives}

\author{James Carmer}
\affiliation{McKetta Department of Chemical Engineering, University of Texas at Austin, Austin, Texas 78712, USA}

\author{Avni Jain}
\affiliation{McKetta Department of Chemical Engineering, University of Texas at Austin, Austin, Texas 78712, USA}

\author{Jonathan A. Bollinger}
\affiliation{McKetta Department of Chemical Engineering, University of Texas at Austin, Austin, Texas 78712, USA}

\author{Frank van Swol}%
\affiliation{Sandia National Laboratories, Department 1814, P.O. Box 5800, Albuquerque, NM 87185, USA}

\author{Thomas M. Truskett}
\email{truskett@che.utexas.edu}
\affiliation{McKetta Department of Chemical Engineering, University of Texas at Austin, Austin, Texas 78712, USA}

\date{\today}

\begin{abstract}
Molecular dynamics simulations and a stochastic Fokker-Planck equation based approach are used to illuminate how position-dependent solvent mobility near one or more tracer particle(s) is affected when tracer-solvent interactions are rationally modified to affect corresponding solvation structure.
For tracers in a dense hard-sphere fluid, we compare two types of tracer-solvent interactions: 
(1) a hard-sphere-like interaction; and 
(2) a soft repulsion extending beyond the hard core designed via statistical mechanical theory to 
enhance tracer mobility at infinite dilution by suppressing coordination-shell structure (Carmer et al., \emph{Soft Matter} \textbf{2011}, \emph{8}, 4083). 
For the latter case, we show that the mobility of surrounding solvent particles is also increased by addition of the soft repulsive interaction, which helps to rationalize the mechanism underlying the tracer's enhanced diffusivity. 
However, if multiple tracer surfaces are in closer proximity (as at higher tracer concentrations), similar interactions that disrupt local solvation structure instead suppress the position-dependent solvent dynamics. 

\end{abstract}

\pacs{Valid PACS appear here}
\maketitle

\section{Introduction}

Engineering the transport properties of colloidal and nanoparticle additives in dispersions is of great fundamental interest and has practical implications in a wide range of technologies such as drug-delivery mechanisms, polymer nanocomposites (PNCs), material fabrication techniques, and separations.~\cite{saltzman2001drug,wasan2003spreading,PRL2014SKumar} A useful feature of such dispersions is that their transport behavior can often be systematically varied via tuning the effective inter-species interactions, e.g., external electric fields have been used to control the dynamics of conducting nanoparticles~\cite{MBevan2014}, solvent pH has been tuned to manipulate the drug-release kinetics of nanocapsules~\cite{MoraHuertas2010113}, etc.

One approach to rationally tune transport properties is to utilize static-dynamic correlations, such as the empirical and quasi-universal positive correlation 
between excess entropy and long-time particle mobility (e.g., diffusivity) for bulk fluids  ~\cite{Rosenfeld1977,Dzugutov1996,Rosenfeld1999,Hoyt2000,Li2005,Sharma06,Mittal2006s2,GnanDyre2009,Krekelberg2009,PondJCP2011,Ingebrigtsen2012,Nayar13,Dyre2014} and various inhomogeneous fluids~\cite{Mittal2006,Mittal2007,Mittal2007mixtures,GGPRL2008,Goel2009,Chopra2010confined,Borah12,Ma2013,Liu2013,Ingebrigtsen2013}.
Another is through approximate theoretical approaches, like mode-coupling theory (MCT), which connect static structure and dynamic relaxation properties~\cite{JKGDhont1998JCP,Krakoviack2007,KSSchweizer2008,Lang2010,Lang2014}.
Within such a framework, one can use statistical mechanical theories like fundamental measure theory (FMT)~\cite{Roth2010} to connect interparticle interactions and structural correlations, the latter of which are then approximately related to the transport properties of the system. Using these concepts, one can engineer interparticle interactions for targeted dynamic properties.

In this vein, previous work using stochastic optimization and statistical mechanical theory found that the mobility of a single tracer particle in a dense fluid could be enhanced (relative to a hard-sphere-like tracer) by adopting a ``flattening'' tracer-solvent interaction, which interestingly also increased the effective tracer diameter~\cite{Carmer2012}. In particular, the flattening interactions were designed to maximize the entropy associated with the tracer's static interparticle correlations, and hence disrupt the coordination shells of the surrounding solvent particles~\cite{Carmer2012}. However, several interesting questions emerge from these findings that remain unresolved, including: why should a ``larger'' tracer exhibit faster dynamics? how are the dynamics of the \emph{solvent} particles influenced by tuning coordination structure? and what happens to the solvent (and tracer) dynamics if multiple tracers are in close proximity, i.e., the average tracer concentration is increased?

In this paper, we take steps to address these questions by using a recently introduced steady-state ``color'' reaction-counterdiffusion particle labeling approach~\cite{Carmer2014} to calculate position-dependent diffusivities of the solvent particles surrounding the tracer additive. This method allows us to isolate the affect of the modified tracer on the dynamic behavior of the surrounding fluid relative to the hard-sphere like tracer, and provides mechanistic insights into the underlying phenomenon of how disordering of the surrounding fluid affects the tracer-fluid system.

Furthermore, we explore the impact of tracer concentration on the local coordination-shell dynamics for the modified and hard-sphere-like tracer particle, by approximating the solvent-tracer system in a thin-film morphology, where the solvent particles are confined by two planar surfaces (representing the tracer particle surface, Fig.~\ref{sch:Figure1}). This model is motivated by experimental studies that have also shown the utility of approximating the bulk material properties of high-additive composites via examination of thin films having similar surface properties.~\cite{rittigstein2007model,ramanathan2008functionalized}.

\section{Computational Methods}

\begin{figure}
  \includegraphics[width=0.4\textwidth]{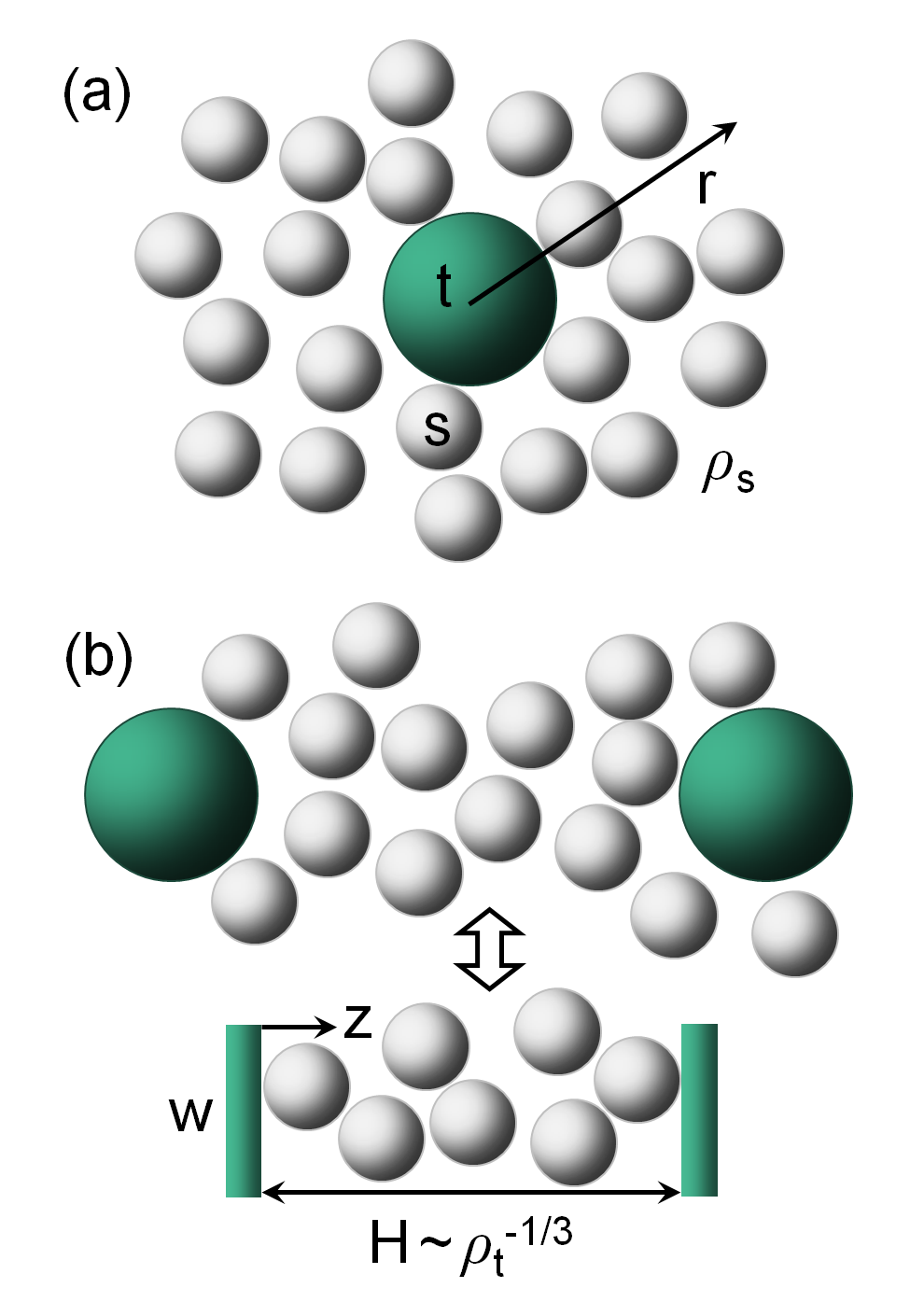}
  \caption{(color online). Schematics illustrating the two system geometries under consideration. (a) Tracer particle (t) at infinite dilution surrounded by a 3D volume of dense solvent (s) particles, where we consider solvent motions along the radial \(r\)-coordinate relative to the tracer center of mass. (b) Thin film approximation of systems with higher tracer concentration \(\rho_{\text{t}}\), where solvent particles are confined between walls (w) representing tracer surfaces. The walls are separated by thickness \(H\) in the \(z\)-direction, and periodic boundaries are employed in the \(x\)- and \(y\)-directions. Particles in (a) and (b) are drawn on the same scale.}
  \label{sch:Figure1}
\end{figure}

For both the single-tracer and thin film systems shown in Fig.~\ref{sch:Figure1}, we generate particle trajectories from 3D molecular dynamics (MD) simulations in the canonical ensemble, where trajectories are evolved using the velocity-Verlet method~\cite{Rapaport2004}. We use a time-step of \(0.001 \sigma_{\text{s}}\sqrt{m/\epsilon}\), where \(\sigma_{\text{s}}\), \(m\), and \(\epsilon\) are the characteristic solvent (s) diameter, mass, and energy scales, respectively. The temperature is constrained to \(T=\epsilon/k_{\rm B}\) (where \(k_{B}\) is Boltzmann's constant) using the Nos\'{e}-Hoover thermostat.

For the case in Fig.~\ref{sch:Figure1}(a), we incorporate a single tracer (t) with diameter \(\sigma_{t}\) into a fluid comprised of \(N_{\text{s}} = 8000\) hard-sphere (HS) solvent particles. Solvent-solvent HS interactions are approximated by the following continuous, steeply-repulsive Weeks-Chandler-Andersen (WCA) pair potential~\cite{ChandlerWeeksAndersenScience1983}: \(\varphi_{\text{ss}}(r) = \varphi^{\text{WCA}}(r,\sigma_{\text{s}},\epsilon) = 4\epsilon([\sigma_{\text{s}}/r]^{48} - [\sigma_{\text{s}}/r]^{24})+\epsilon\) for \(r\leq2^{1/24}\sigma_{\text{s}}\) and \(\varphi_{\text{ss}}(r)=0\) for \(r>2^{1/24}\sigma_{\text{s}}\), where \(r\) is the interparticle separation. The tracer-solvent pair interaction is defined by: \(\varphi_{\text{ts}}(r) = \varphi^{\text{WCA}}(r,\sigma_{\text{ts}},\epsilon)\) for \(r < r_0\) and \(\varphi_{\text{ts}}(r) = \varphi^{\text{WCA}}(r,\sigma_{\text{ts}},\epsilon) + \varphi_{\text{ts},0}(r)\) for \(r \geq r_0\), where \(\sigma_{\text{ts}} = (\sigma_{\text{t}}+\sigma_{\text{s}})/2\). Here, \(\varphi_{\text{ts},0}(r)\) is a contribution outside the hard core that can be tuned to affect the solvent density profiles around the tracer, and \(r_0\) is chosen such that the tuning procedure does not alter the range of the hard-core exclusion interaction.~\cite{Carmer2012}

For the thin films in Fig.~\ref{sch:Figure1}(b), we simulate \(N_{\text{s}}=4000\) solvent particles confined between between two flat walls (w) in the \(z\)-direction separated by a thickness \(H\), and are periodic in the \(x\)- and \(y\)-directions. The wall-solvent (or tracer-solvent) interaction is defined by \(\varphi_{\text{ws}}(z) = \varphi^{\text{WCA}}(z+0.5,\sigma_{\text{ss}},\epsilon)\) for \(z < z_0\)
and \(\varphi_{\text{ws}}(z) = \varphi^{\text{WCA}}(z+0.5,\sigma_{\text{ss}},\epsilon) + \varphi_{\text{ws},0}(z)\) for \(z \geq z_0\). This term is equivalent to the solvent-solvent hard-core interaction combined with an
additional variable \(\varphi_{\text{ws},0}(z)\) that can be tuned to affect the solvent density variations across the film. Here, we define \(\varphi_{\text{ws}}(z)\) as a function of \((z+0.5)\) such that solvent centers can access positions ranging from approximately \(0.5 \leq z \leq (H-0.5)\), as is the case for true hard spheres situated between flat hard walls.

We are primarily interested in measuring how solvent motions near the tracer and wall surfaces depend upon the type of density profiles allowed in the near-surface solvation layers. Notably, we are probing diffusive solvent displacements along paths that are inhomogeneous--i.e., motions along the \(r\)- and \(z\)-coordinates that are subject to non-isotropic solvent density fields--such that we cannot estimate the corresponding position-dependent solvent diffusivities from particle displacements using the typical Einstein relation~\cite{LiuBerne2004}. (However, the Einstein relation is applied to obtain, e.g., \emph{isotropic} tracer diffusivity \(D_{\text{t}} = \langle \Delta \mathbf{r}^2 \rangle / 6\Delta t\), where \(\langle \Delta \mathbf{r}^2 \rangle\) is the tracer mean squared displacement in the \(x\), \(y\), and \(z\) directions over lag times \(\Delta t\) exceeding the timescales of ballistic motions.)

Position-dependent particle diffusivities in dense inhomogeneous fluids are accurately described by the Fokker-Planck (FP) equation~\cite{JMPRL2008,JMJCP2012,Bollinger2014,Carmer2014}. For the single-tracer case in Fig. 1(a), the FP equation describing solvent displacements along the \(r\)-coordinate of the (reference) tracer particle is~\cite{JMPRL2008}
\begin{equation}
\label{eq:FP}
\dfrac{\partial G}{\partial t} = \dfrac{1}{r^{\text{2}}}\dfrac{\partial}{\partial r} \Bigg[ r^{\text{2}}D^{*}_{\text{s,r}}(r)\Bigg(\beta \dfrac{dV_{\text{ts}}}{dr}G + \dfrac{\partial G}{\partial r}\Bigg)\Bigg]
\end{equation}
which contains position-dependent diffusivities in the \(r\)-direction \(D^{*}_{\text{s,r}}(r)\). Here, \(G(r, t_0+\Delta t|r', t_0)\) is the Markovian propagator describing temporal single-particle displacements given a non-uniform potential of mean force (PMF). For the solvent surrounding a single tracer, this PMF is given by \(V_{\text{ts}}(r)=-\ln\{[\rho_{\text{ts}}/\rho_{\text{ts,avg}}](r)\} + C\), where \(\rho_{\text{ts}}(r)\) is the solvent density at some distance \(r\); \([\rho_{\text{ts}}/\rho_{\text{ts,avg}}](r)\) is the partial radial distribution function (i.e., \(g_{\text{ts}}(r)\)); and \(C\) is an arbitrary constant.

To extract \(D_{\text{s,r}}(r)\) profiles from particle trajectories, we use a color reaction-counterdiffusion treatment of the steady-state form of eq.~\ref{eq:FP} (i.e., \(\partial G/\partial t=0\))~\cite{Carmer2014}. We then rescale the \(D_{\text{s,r}}(r)\) profiles that account for the relative mobilities of the the tracer particles themselves, where \(D_{\text{s,r}}(r) = D^{*}_{\text{s,r}}(r) \left[ D^{\text{blk}}_{\text{s}} / ( D^{\text{blk}}_{\text{s}}+D_{\text{t}} ) \right]\), \(D^{*}_{\text{s,r}}(r)\) are the non-normalized local diffusivities obtained from eq.~\ref{eq:FP}, \(D^{\text{blk}}_{\text{s}}\) is the solvent bulk diffusivity in the absence of any tracer, and \(D_{\text{t}}\) is the long-time tracer diffusivity. Thus, \(D_{\text{s,r}}(r)\) approaches \(D^{\text{blk}}_{\text{s}}\) as \(r \rightarrow \infty\) independent of the type of tracer-solvent interaction. Implementation details and the analogous expressions for local solvent diffusivities along the \(z\)-direction \(D_{\text{s,z}}(z)\) for the thin film systems can be found in a previous publication~\cite{Carmer2014}.

\section{Results \& Discussion}

\begin{figure}
  \includegraphics{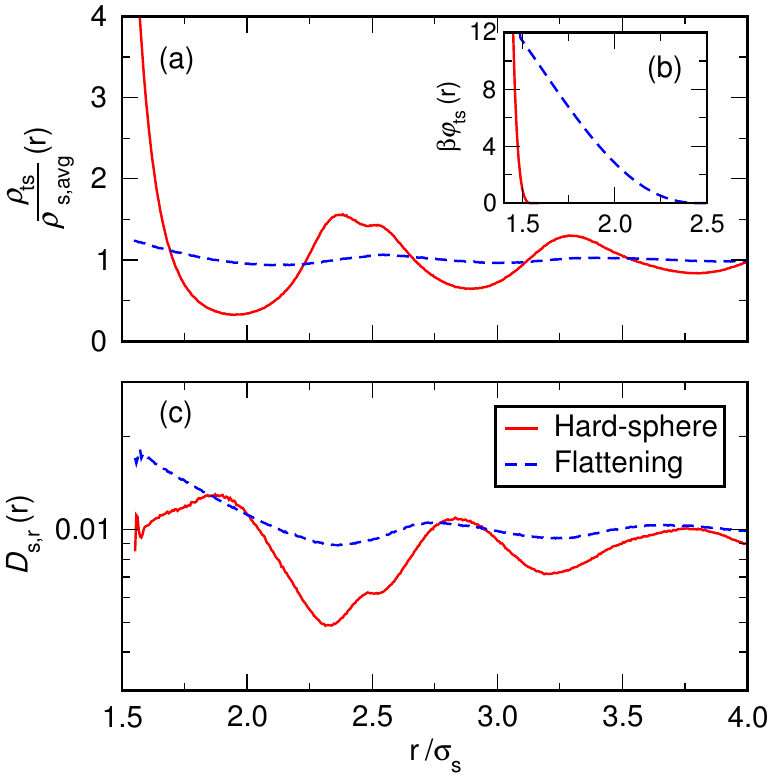}
  \caption{(color online). (a) Radial density profiles of solvent particles \([\rho_{\text{s}}/\rho_{\text{s,avg}}](r)\) surrounding tracer particles of diameter \(\sigma_{t}/\sigma_{\text{s}}=2.0\) for the two different tracer-fluid interactions \(\varphi_{\text{ts}}(r)\) 
shown in inset (b). Red solid `hard-sphere' profiles correspond to bare WCA potential for \(\varphi_{\text{ts}}(r)\), while blue dashed `flattening' profiles correspond to a \(\varphi_{\text{ts}}(r)\) optimized to flatten \([\rho_{\text{s}}/\rho_{\text{s,avg}}](r)\) and thus maximize the contribution of the tracer particle's correlations to the two-body excess entropy \({s}^{(2)}_{\text{ts}}\) associated with tracer-solvent static correlations. (c) Corresponding position-dependent diffusivities of fluid particles \(D_{\text{s,r}}(r)\) along the radial direction.}
  \label{sch:Figure2}
\end{figure}

We begin our discussion by considering Fig.~\ref{sch:Figure2}, where we demonstrate how solvent structure and dynamics around infinitely dilute tracer particles are affected by two different types of tracer-solvent interactions \(\varphi_{\text{ts}}\) shown in Fig. 2(b): (1) a hard-sphere-like steep WCA repulsion (i.e., the contribution outside the hard core \(\varphi_{t,0}(r)=0\) for \(r \geq r_0\)); and (2) the same core repulsion combined with a softer long-range repulsion (resembling Yukawa screened electrostatic interactions)
to flatten (i.e., eliminate) solvent coordination shells around the tracer. As shown previously~\cite{Carmer2012} for a single tracer, adopting such flattening potentials minimizes tracer-solvent structural pair correlations (\(g_{\text{ts}}\)), which increases the corresponding two-body excess entropy \(s_{\text{2,ts}}\) 
relative to the hard-sphere-like tracer case, as evident from its definition: 
\begin{equation} \label{eq:s2bulk}
s_{\text{2,ts}}/k_{\text{B}} = -\dfrac{\rho_{\text{s}}}{2} \int_{0}^{\infty} \{ g_{\text{ts}}(r) \ln{g_{\text{ts}}(r)} - g_{\text{ts}}(r) + 1 \}\mathrm{d}\vec{r}
\end{equation}
\noindent 
In turn, the long-time tracer diffusivity \(D_{\text{t}}\) can be enhanced by a factor of up to two or more compared to the hard-sphere-like WCA case depending on the size ratio of the tracer and solvent particles \(\sigma_{\text{t}}/\sigma_{\text{s}}\).~\cite{Carmer2012} This is--at first glance--counterintuitive because the tracers with the flattening potentials have larger apparent diameters, which one might na\"ively expect to \emph{depress} diffusive mobility.

In Fig.~\ref{sch:Figure2}(a) and ~\ref{sch:Figure2}(c), we show the position-dependent solvent densities \([\rho_{\text{s}}/\rho_{\text{s,avg}}](r)\) (i.e., \(g_{\text{ts}}(r)\)) and  diffusivities \(D_{\text{s,r}}(r)\) around these two tracer particles of diameter \(\sigma_{\text{t}}/\sigma_{\text{s}}=2.0\). 
The average tracer diffusivities from both cases were related as: \(D^{\text{flat}}_{\text{t}} / D^{\text{WCA}}_{\text{t}} \simeq 2\) (see Fig. 3 in ref.~\cite{Carmer2012}). Here, it is evident that using the flattening \(\varphi_{\text{ts,0}}(r)\) thoroughly destroys solvent coordination shells while simultaneously enhancing local solvent diffusivities \(D_{\text{s,r}}(r)\) at virtually all distances at and near the tracer surface, which helps to rationalize the corresponding enhancement in tracer diffusivity.
The solvent diffusivity measurements for the two cases reflect known excess entropy scalings for transport coefficients and support the notion that tracer and solvent dynamics are coupled, i.e., the shift in solvent dynamics drives (even non-intuitive) trends in relative tracer mobility.

\begin{figure}
  \includegraphics{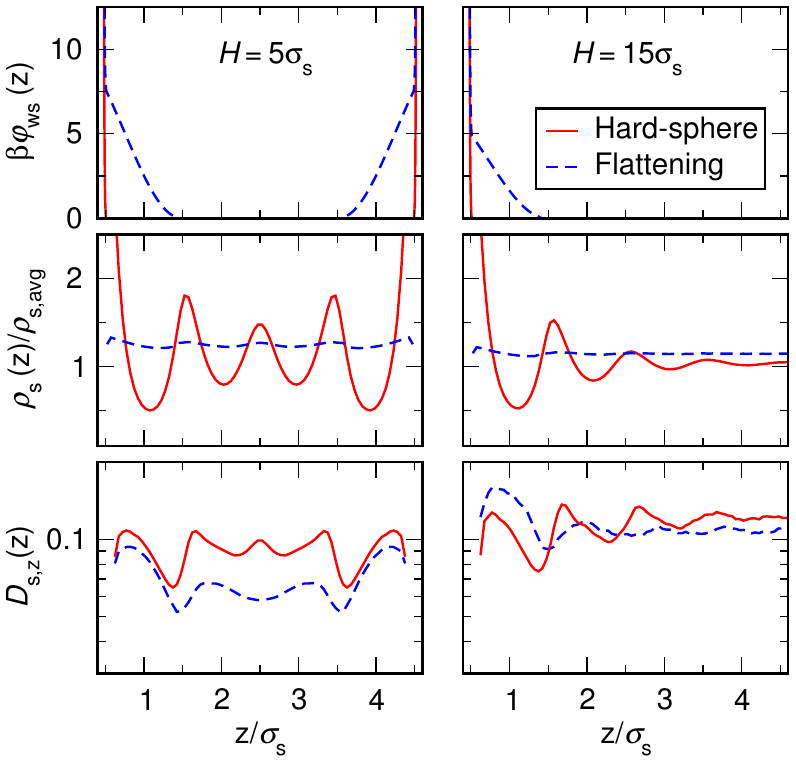}
  \caption{(color online) Results for thin film systems: (top) wall-solvent interactions \(\beta\varphi_{\text{ws}}(z)\); (middle) solvent density profiles \(\rho_{\text{s}}(z)\) normalized by average density \(\rho_{\text{s,avg}}\); and (bottom) position-dependent diffusivities of solvent particles \(D_{\text{s,z}}(z)\) in the \(z\)-direction. Left panels correspond to film thickness \(H=5\) and right panels to \(H=15\), where in both cases the average solvent packing fraction is \(\phi=0.35\) (based on total channel thickness \(H\)). Red solid `hard-sphere' profiles correspond to the WCA potential for \(\beta\varphi_{\text{ws}}(z)\), while blue dashed `flattening' profiles correspond to \(\beta\varphi_{\text{ws}}(z)\) potentials derived via FMT to flatten \(\rho_{\text{s}}(z)\) across \(H\).}
  \label{sch:Figure3}
\end{figure}

While the discussion above shows that destroying coordination shells enhances tracer and solvent dynamics for infinitely dilute tracers, a natural question is whether the use of this strategy would have the same implications for systems with finite tracer concentrations \(\rho_{\text{t}} > 0\). To address this, we consider how solvent dynamics are affected in the thin film systems illustrated in Fig.~\ref{sch:Figure1}(b), where the interwall separation \(H\) approximately scales with tracer concentration according to \(H \propto \rho_{\text{t}}^{1/3}\). In this way, we examine whether destroying density variations \emph{categorically} enhances nearby solvent diffusivity--and tracer diffusivity, due to their coupling--regardless of the average distance between proximal ``tracer'' surfaces. 
\footnote{Measurements of solvent dynamics in proximity to multiple surfaces are made significantly more accessible due to the adoption of the Cartesian geometry (i.e., we avoid considering solvent motions relative to two moving curved surfaces). This is also advantageous because it is difficult to unambiguously derive tracer-tracer interactions \(\varphi_{\text{tt}}(r)\) given target tracer-solvent structure \(g_{\text{ts}}(r)\), making the explicit simulation of multiple tracers quite challenging.}

In Fig.~\ref{sch:Figure3}, we show the position-dependent solvent density \(\rho_{\text{s}}(z)\) and diffusivity \(D_{\text{s,z}}(z)\) profiles associated with wall-solvent interactions \(\varphi_{\text{ws}}(z)\) that are either steeply repulsive (i.e., \(\varphi_{\text{w,0}}(z)=0\)) or have been augmented with \(\varphi_{\text{w,0}}(z)\) potentials derived from Fundamental measure theory (FMT)~\cite{GGPRL2008,Roth2010} to flatten the solvent density profiles \(\rho_{\text{s}}(z)\) across different film thicknesses \(H\). Based on the top panels of Fig.~\ref{sch:Figure3}, it is evident that ramp-like \(\varphi_{\text{w,0}}(z)\) repulsions spanning the characteristic particle lengthscale \(\sigma_{\text{s}}\) are effective at destroying near-surface density variations. Such flattening interactions also resemble the \(\varphi_{\text{t,0}}(z)\) flattening interactions for spherical tracers. By examining the middle- and bottom-left panels of Fig.~\ref{sch:Figure3}, one can see qualitatively different dynamic responses to flattening \(\rho_{\text{s}}(z)\) for various \(H\): the diffusivities \(D_{\text{s,z}}(z)\) are \emph{uniformly depressed} for the highly confined \(H=5\) case while they are instead \emph{enhanced} near the walls for \(H=15\). The latter observation is expected as \(H \rightarrow \infty\) qualitatively corresponds to the single-tracer limit \(\rho_{\text{t}} \rightarrow 0\).

We can understand these opposing results in a general way by considering particle packing effects within thin films. In the case of unmodified hard-wall-like WCA boundaries, particles tend to accumulate near the walls to minimize excluded volume. Given that the external potentials capable of flattening \(\rho_{\text{s}}(z)\) are soft repulsions that ``push'' these particles toward the center of the film, it is perhaps unsurprising that for very thin films, there is insufficient space for these particles to redistribute themselves in a way that allows for efficient packing and corresponding diffusive motions~\cite{GGPRL2008}. In contrast, as the thickness of the thin film increases, particles near the walls constitute a smaller fraction of the total fluid population. Thus, their redistribution away from the walls is accomodated more readily, which should not unconditionally frustrate near-surface diffusion.

\begin{figure}
  \includegraphics{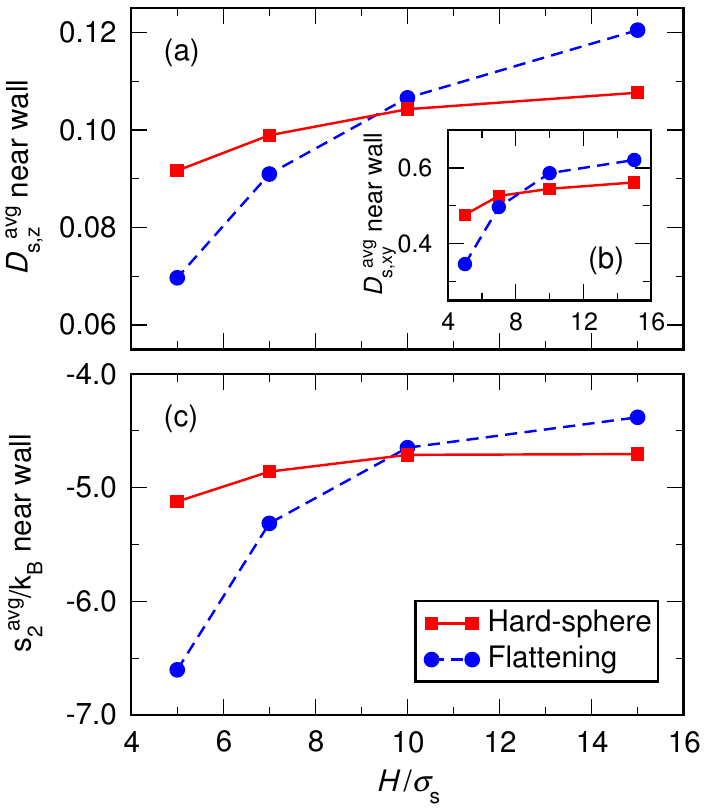} 
  \caption{(color online) (a) Average fluid diffusivities in the \(z\)-direction \(D_{\text{s,z}}^{\text{avg}}(z)\) for various thin film thicknesses \(H/\sigma_{\text{s}}\), where \(D_{\text{s,z}}^{\text{avg}}(z)\) is calculated based on \(z\)-positions near the confining surfaces (see text). (b) Average diffusivities in the unconfined directions \(D_{\text{xy}}\) for fluid particles within the same regions. (c) Average two-body excess entropies \(s_{2}^{\text{avg}}\) quantifying static correlations between solvent particles within the same regions. Lines connecting symbols are a guide to the eye.}
  \label{sch:Figure4}
\end{figure}

In Fig.~\ref{sch:Figure4}, we demonstrate that this dichotomy in diffusive responses to flattening \(\rho_{\text{s}}(z)\) at various \(H\) is systematic and can be rationalized by considering the corresponding changes in the static solvent-solvent correlations, as quantified by excess entropy. To make this analysis consistent, we calculate \emph{average} solvent diffusivities both perpendicular \(D_{\text{s,z}}^{\text{avg}}\) and lateral \(D_{\text{s,xy}}^{\text{avg}}\) to the hard-sphere-like and flattening potentials for particles located in regions near the thin film walls. These regions span \(0 \leq z \leq d_{\text{ws}}\) and \((H-d_{\text{ws}}) \leq z \leq H\), where \(d_{\text{ws}} = 2.5\sigma_{\text{s}}\) for all values of \(H\) (this \(d_{\text{ws}}\) corresponds to complete coverage for the \(H/\sigma_{\text{s}}=5\) film). We also calculate average two-body excess entropies \(s_2^{\text{avg}}\) using the same particle trajectories within this region.

Given \(D_{\text{s,z}}(z)\) diffusivity profiles, average diffusivities in the near-surface regions are calculated via
\begin{equation} \label{eq:Davg}
D^{\text{avg}}_{\text{s,z}} = \sum_{i=0,1} \dfrac{1}{2}\int_{z_{i}}^{z'_{i}}D_{\text{s,z}}(z)\rho(z)\mathrm{d}z\text{ }/\int_{z_{i}}^{z'_{i}}\rho(z)\mathrm{d}z
\end{equation}
\noindent 
where \(z_0 = 0\), \(z'_0 = d_{\text{ws}}\), \(z_1 = H-d_{\text{ws}}\), and \(z'_1 = H\). 
We calculate lateral diffusivities via the Einstein relation \(D_{\text{s,xy}}^{\text{avg}} = \langle \Delta \mathbf{r}^2 \rangle / 4\Delta t\) using mean squared displacements \(\langle \Delta \mathbf{r}^2 \rangle\) in the periodic \(x\) and \(y\) directions. Displacements are aggregated from any particles that are located within \(d_{\text{ws}}\) of the nearest film boundary at time \(t\) up until the maximum time lag \(\Delta t_{\text{max}} = t'-t\), where \(t'\) is the time at which the particles exit the near-wall region.
\footnote{For \(H=5\), this corresponds to tracking all fluid particles for \(\Delta t \rightarrow \infty\). For larger pores, one achieves good statistics for \(\Delta t \leq 100\), since beyond this lag time most particles have ``escaped'' the near-wall region. Nonetheless, this \(\Delta t\) is much greater than the characteristic time \(\tau_{\text{C}}\) at which particles motions become diffusive, and allows for straightforward linear fits of the displacements to obtain \(D_{xy}\).}
We obtain \(s_2^{\text{avg}}\) values via an integral analogous to eq.~\ref{eq:Davg} over position-dependent two-body excess entropy \(s_2(z)\), where \(s_2(z)\) profiles are calculated using a recast form of eq.~\ref{eq:s2bulk} in the Cartesian geometry~\cite{Bollinger2014}.

In Fig.~\ref{sch:Figure4}(a-b), we compare average solvent diffusivities measured near surfaces with `hard-sphere-like' and `flattening' \(\varphi_{\text{ws}}(z)\) interactions for various \(H\), where it is evident that the effect of flattening density variations upon solvent mobility is \emph{qualitatively} dependent upon inter-surface proximity. For \(H \sigma_{\text{s}} \lesssim 10\), average dynamics in all directions are slowed upon flattening the density profile, while for \(H/\sigma_{\text{s}} \gtrsim 10\), dynamics are instead enhanced (consistent with the tracer case). This crossover can be unified with the single-tracer case by considering the \(s^{\text{avg}}_2\) curves shown in Fig.~\ref{sch:Figure4}(c), which illustrate that the relative slowdown for \(H/\sigma_{\text{s}} \lesssim 10\) reflects the more general positive correlation between particle mobility and multi-body excess entropy. Thus, eliminating one-body density variations has a non-trivial \(H\)-dependent effect upon two-body static correlations, where the latter are more meaningfully correlated with dynamics. (For another pronounced example of such effects, see Goel et al.~\cite{GGPRL2008}.)

Crucially, the trends in Fig.~\ref{sch:Figure4} imply that tracer-solvent interactions \(\varphi_{\text{ts}}(r)\) designed at infinite dilution to enhance tracer (and solvent) mobility may not generally have the same qualitative impact at sufficiently high tracer concentrations (i.e., sufficiently thin inter-tracer solvent regions). Using the crossover film thickness \(H^{*}/\sigma_{\text{s}} \simeq 10\) from Fig.~\ref{sch:Figure4}, we can obtain an order-of-magnitude estimate for the limiting tracer concentration \(\rho_{\text{t}}^{*}\) beyond which the single-tracer physics might be expected break down due to solvent packing effects: our most conservative (i.e., biased towards a greater value of \(\rho_{\text{t}}^{*}\)) calculations indicate that \(\rho_{\text{t}}^{*} \approx \mathcal{O}(0.01)\).
\footnote{We noted before that the thin film thickness \(H\) scales with tracer particle density \(\rho_{\text{t}}\) as \(H \sim {\rho_{\text{t}}}^{\text{-1/3}}\).
For a close-packed FCC tracer lattice, which corresponds to the densest tracer arrangement for a given nearest-neighbor intersurface distance \(H\), \(H/\sigma_{\text{t}}=2^{1/6}{\rho_{\text{t}}}^{-1/3} -1\) (note that here the characteristic lengthscale is \(\sigma_{\text{t}}\)). From Fig.~\ref{sch:Figure4}, the crossover intersurface distance is \(H^{*}/\sigma_{\text{s}}=10\), which corresponds to \(H^{*}/\sigma_{\text{t}}=5\) because \(\sigma_{\text{t}}/\sigma_{\text{s}}=2\). One can then calculate the crossover tracer density by using \(\rho_{\text{t}}^{*}=\sqrt{2}/(H^{*}/\sigma_{\text{t}}+1)^3=0.0065\).
Thus, for tracer densities \(\rho_{\text{t}} \geq 0.0065\), packing effects upon flattening the density profile would suppress solvent mobility.} 
That \(\rho_{\text{t}}^{*}\) is so low lends a cautionary note in terms of deploying tracer-solvent interactions designed in the dilute limit at any significantly higher concentrations.

\section{Conclusions}

Using recently introduced techniques for characterizing the position-dependent dynamics of inhomogeneous fluids, we have shown that eliminating the coordination structure of an infinitely dilute tracer (additive) in bulk solvent to decrease static correlations (increase excess entropy)--which is achieved by rationally tuning the tracer-solvent pair interaction--enhances the diffusive mobilities of both the tracer and the surrounding solvent particles. However, upon incorporating similarly tuned interactions into thin films of solvent particles, which approximate systems of higher tracer concentration, we find that eliminating one-body solvation structure for film thicknesses (i.e., tracer-tracer distances) smaller than several solvent particle diameters decreases excess entropy and suppresses solvent mobility. This observation nicely explains the results of two previous studies where tuning interactions to increase excess entropy had opposite effects on the dynamics of solvent surrounding a tracer particle~\cite{Carmer2012} or trapped between interacting surfaces~\cite{GGPRL2008}. Due to the apparent coupling of tracer and solvent dynamics, this is suggestive that tracer-solvent interactions designed at dilute conditions could have qualitatively different impacts upon system dynamics at sufficiently high tracer concentrations, where the critical loadings separating these regimes are likely very low.

\section{Acknowledgments}

This work was supported by the Gulf of Mexico Research Initiative, the Robert A. Welch Foundation (F-1696) and the National Science Foundation (CBET-1403768). FVS acknowledges support by the United States Department of Energy, Office of Basic Energy Sciences, Division of Materials Sciences and Engineering and Sandia's LDRD program. Sandia National Laboratories is a multi-program laboratory managed and operated by Sandia Corporation, a wholly owned subsidiary of Lockheed Martin Corporation, for the U.S. Department of Energy's National Nuclear Security Administration under contract DE-AC04-94AL85000. We also acknowledge the Texas Advanced Computing Center (TACC) at The University of Texas at Austin for providing HPC resources for this study.

%

\end{document}